\documentclass[a4paper]{article}
\usepackage{amscd,amsmath,amssymb}

\newcommand{\bb}[1]{ \boldsymbol{#1}{}}

\newcommand{\p}[1]{(\ref{#1})}

\newcommand{\cL}{{\cal L}}

\newcommand{\bD}{{\overline D}{}}

\newcommand{\bQ}{{\overline Q}{}}
\newcommand{\bS}{{\overline S}{}}

\newcommand{\bI}{{\overline I}{}}
\newcommand{\bB}{{\overline B}{}}
\newcommand{\bC}{{\overline C}{}}

\newcommand{\bpsi}{{\bar\psi}{}}
\newcommand{\brho}{{\bar\rho}{}}

\newcommand{\bv}{{\bar v}}

\newcommand{\cN}{{ {\cal N}   }}

\newcommand{\und}{\qquad\textrm{and}\qquad}

\newcommand{\be}{\begin{equation}}
\newcommand{\ee}{\end{equation}}
\newcommand{\bea}{\begin{eqnarray}}
\newcommand{\eea}{\end{eqnarray}}

\newcommand{\ba}{\begin{array}} \newcommand{\ea}{\end{array}}

\def\im{{\rm i}}
\def\diff{{\rm d}}

\newcommand{\nn}{\nonumber}

\begin{document}
\begin{flushright}
\end{flushright}

\vspace{1cm}

\begin{center}
{\LARGE\bf Supersymmetric many-body Euler-Calogero-Moser model}

\end{center}
\vspace{1cm}

\begin{center}
{\large\bf  Sergey Krivonos${}^{a,b}$, Olaf Lechtenfeld$^c$ and Anton Sutulin${}^a$}
\end{center}

\vspace{0.2cm}

\begin{center}
{${}^a$ \it
Bogoliubov  Laboratory of Theoretical Physics, JINR,
141980 Dubna, Russia}

{${}^b$ \it St. Petersburg Department of
V.A. Steklov Institute of Mathematics of
the Russian Academy of Sciences,\\
27 Fontanka, St. Petersburg, Russia}

${}^c$ {\it
Institut f\"ur Theoretische Physik and Riemann Center for Geometry and Physics, \\
Leibniz Universit\"at Hannover,
Appelstrasse 2, D-30167 Hannover, Germany}

\vspace{0.5cm}

{\tt krivonos@theor.jinr.ru, lechtenf@itp.uni-hannover.de, sutulin@theor.jinr.ru}
\end{center}
\vspace{2cm}

\begin{abstract}\noindent
We explicitly construct a supersymmetric $so(n)$ spin-Calogero model
with an arbitrary even number $\cal N$ of supersymmetries.
It features $\frac{1}{2}{\cal N}n(n{+}1)$ rather than ${\cal N}n$ fermionic coordinates
and a very simple structure of the supercharges and the Hamiltonian.
The latter, together with additional conserved currents, form an $osp({\cal N}|2)$ superalgebra.
We provide a superspace description for the simplest case, namely ${\cal N}{=}2$ supersymmetry.
The reduction to an $\cal N$-extended supersymmetric goldfish model is also discussed.
\end{abstract}


\vskip 0.5cm


\newpage

\setcounter{equation}{0}
\section{Introduction}
In recent years notable progress was achieved in the supersymmetrization of the bosonic matrix models \cite{F1,F2,F3,F4,F5,SUSYCal}.
It has been known for a long time that matrix models are an efficient tool of constructing conformally
invariant systems (see e.g.~\cite{Poly1} and refs.\ therein) For example, the Calogero model as well as its different extensions \cite{sCal,GH,Poly2,AH1,gf}
are closely related to matrix models and can be obtained from them by a reduction procedure. The supersymmetrization of matrix models consists
in replacing the bosonic matrix entries by superfields \cite{F1,F2,F3,F4,F5}. While this approach has been quite successful for $\cN \leq 4$ extended supersymmetry,
it seems to be less efficient or even inapplicable for $\cN >4$ supersymmetric cases.\footnote{
An up to now unique example of a matrix system with $\cN=8$ supersymmetry has appeared in \cite{F5} in $\cN=4$ superspace.}
In contrast, the Hamiltonian approach has no serious restriction on the number of supersymmetries, due to the absence of auxiliary components.

The key feature of a supersymmetric extension of one-dimensional models within the Hamiltonian approach is the appearance of additional fermionic matrix
degrees of freedom accompanying the standard $\cN\,n$ fermions customarily required for an $\cN$-extended supersymmetric system with $n$~bosonic coordinates.
Recently we implemented this feature to construct a supersymmetric extension of Hermitian matrix models which admits an arbitrary number of supersymmetries~\cite{SUSYCal}.
We also provided a supersymmetrization of the reduction procedure which yields an $\cN$-extended $n$-particle supersymmetric Calogero model.
The question we address in this paper is how to (if possible) repeat this supersymmetrization procedure for the real symmetric matrix model~\cite{sCal}.

In the bosonic case, the free matrix model associated with real symmetric matrices (see e.g.~\cite{AH1}) results in
a spin generalization of the $n$-particle Calogero--Moser model, which is also known as the Euler--Calogero--Moser (ECM) model~\cite{sCal,GH} and
described by the Hamiltonian
\be\label{ham0}
H = \frac{1}{2} \sum_{i=1}^n p_i^2 + \frac{1}{2} \sum_{i\neq j}^n\, \frac{\ell_{ij}^2}{\left( x_i-x_j\right)^2}\,.
\ee
It depends on the coordinates $x_i(t)$ and momenta $p_i(t)$ of each particle as well as on the internal degrees of freedom encoded in the angular momenta $\ell_{ij}=-\ell_{ji}$.
The coordinates and momenta satisfy the standard Poisson brackets
\be\label{PB1}
\big\{ x_i, p_j \big\} = \delta_{ij},
\ee
while the Poisson brackets of the angular momenta form the $so(n)$ algebra
\be\label{PB12}
\big\{ \ell_{ij}, \ell_{km} \big\}= \frac{1}{2} \big(\delta_{ik} \ell_{jm}+\delta_{jm} \ell_{ik} -\delta_{jk} \ell_{im}-\delta_{im} \ell_{jk}\big).
\ee
The ECM model with the Hamiltonian \p{ham0} possesses conformal invariance. Indeed, if we define the conserved currents of the dilatation $D$ and conformal boost $K$ as
\be\label{KD}
D =  -\frac{1}{2} \sum_{i=1}^n x_i p_i + t H \qquad\textrm{and}\qquad
K = \frac{1}{2} \sum_{i=1}^n x_i^2 - t \sum_{i=1}^n x_i p_i +t^2 H,
\ee
then it is easy to demonstrate that they generate the one-dimensional conformal algebra $so(1,2)$:
\be\label{cf1}
\big\{H, K\big\} = 2 D, \quad \big\{H, D \big\} = H, \quad \big\{ K,D\big\} = -K.
\ee

The equations of motion which follow from the Hamiltonian \p{ham0},
\be\label{eom0}
\ddot{x}_i = 2 \sum_{k\neq i} \frac{\ell_{ik}^2}{(x_i-x_k)^3} \qquad\textrm{and}\qquad
\dot{\ell}_{ij} = -\sum_{k\neq i,j} \ell_{ik}\ell_{kj} \left( \frac{1}{(x_i-x_k)^2}-
\frac{1}{(x_k-x_j)^2}\right),
\ee
consistently reduce to (see e.g. \cite{FC1,AH1,gf})
\be\label{eom1}
\ddot{x}_i = 2 \sum_{j \neq i} \frac{\dot{x}_i \dot{x}_j}{x_i-x_j}
\ee
upon setting
\be\label{ell0}
\ell_{ij} = -\left( x_i-x_j\right) \sqrt{\dot{x}_i \dot{x}_j}.
\ee
This maximally superintegrable system is known as the goldfish model \cite{FC2,FC3}.

In what follows we will construct an $\cN$-extended supersymmetric generalization of the Hamiltonian \p{ham0} and demonstrate
an $Osp(\cN|2)$ invariance of this $\cN=2 M$ supersymmetric ECM model.
We also provide a superfield description for the simplest case of $\cN=2$ supersymmetry.
Finally, we will perform the supersymmetric version of the reduction \p{ell0}, ending up with an $\cN$-extended supersymmetric
goldfish model.

\setcounter{equation}{0}
\section{$\cN$-extended supersymmetric Euler--Calogero--Moser model}

\subsection{Extended super Poincar\'e algebra}

The bosonic ECM model \p{ham0} can be obtained from a free ensemble of real symmetric matrices.
This feature is parallel to the descendence of the $su(n)$ spin-Calogero model~\cite{GH}
from the Hermitian matrix model (for details see~\cite{Poly1}), for which a supersymmetrization has been constructed in~\cite{SUSYCal}.
In full analogy with that case, to construct $\cN$ supercharges $Q^a$ and $\bQ_b$ generating an $\cN{=}\,2M$ superalgebra
\be\label{NSP}
\big\{ Q^a , \bQ_b \big\} = - 2 \im\, \delta^a_b\, H \und \big\{ Q^a, Q^b \big\}=\big\{ \bQ_a, \bQ_b \big\}=0 \quad\textrm{for}\quad a,b=1,2,\ldots M\, ,
\ee
one has to introduce two types of fermions:
\begin{itemize}
\item $\cN\times n$ fermions $\psi^a_i$ and $\bpsi_{i\,a} =\left( \psi^a_i\right)^\dagger$ with $i=1,\ldots, n$.
These fermions can be combined with the bosonic coordinates $x_i(t)$ into $\cN=2 M$ supermultiplets.
\item $\frac{1}{2}\cN\times n(n{-}1)$ additional fermions $\rho^a_{ij}=\rho^a_{ji}$ and $\brho_{ij \,a} =\left(\rho^a_{ij}\right)^\dagger$ subject to $\rho^a_{ii}=\brho_{ii\,a}=0$ (no sum).
\end{itemize}
In total, we thus utilize $\frac{1}{2}\cN n(n{+}1)$ fermions of type $\psi$ and $\rho$, which we demand to obey the following Poisson brackets
\be\label{PB2}
\big\{ \psi^a_i, \bpsi_{j\,b}\big\} = -\im \delta^a_b \delta_{ij}\und
\big\{ \rho^a_{ij}, \brho_{km\,b}\big\}= -\frac{\im}{2}\delta^a_b \big(1- \delta_{ij}\big) \big(1-\delta_{km}\big)
\big( \delta_{ik}\delta_{jm}+\delta_{im}\delta_{jk}\big).
\ee
Using these fermions one can construct the composite objects
\be\label{Pi}
\Pi_{ij}=-\Pi_{ji}=- \im \Bigl[ \bigl( \psi^a_i -\psi^a_j\bigr) \brho_{ij\,a}+\bigl( \bpsi_{i\,a} -\bpsi_{j\,a}\bigr) \rho^a_{ij}
+ \sum_{k=1}^n \left( \rho^a_{ik}\brho_{kj\,a}-\rho^a_{jk}\brho_{ki\,a}\right)\Bigr],
\ee
which satisfy the $so(n)$ Poisson brackets \p{PB12},
\be\label{sonPi}
 \big\{ \Pi_{ij}, \Pi_{km} \big\}= \frac{1}{2} \big( \delta_{ik} \Pi_{jm}+\delta_{jm} \Pi_{ik} -\delta_{jk} \Pi_{im}-\delta_{im} \Pi_{jk}\big),
\ee
and which Poisson-commute with the fermions $\psi$ and $\rho$ as follows,
\bea\label{Picom}
\big\{ \Pi_{ij}, \psi^a_k\big\} & =& \big( \delta_{ik}-\delta_{jk}\big) \rho^a_{ij}, \nn \\
\big\{ \Pi_{ij}, \rho^a_{km}\big\} &=& -\frac{1}{2} \big( 1- \delta_{km}\big)
\Bigl[ \big( \delta_{ik}\delta_{jm}+\delta_{im}\delta_{jk}\big) \big(\psi^a_i-\psi^a_j\big) \\
&& + \big(\delta_{nm}\delta_{jk}+\delta_{kn}\delta_{jm}\big)\rho^a_{in}-
\big(\delta_{nm}\delta_{ik}+\delta_{kn}\delta_{im}\big)\rho^a_{jn}\Bigr]. \nn
\eea

The key idea for constructing the supercharges $Q^a, \bQ_a$ generating \p{NSP} is to ``prolong'' $\ell_{ij}$ to $\ell_{ij}+\Pi_{ij}$ in all expressions, leading to
\be\label{QQb}
Q^a= \sum_{i=1}^n p_i \psi^a_i - \sum_{i \neq j}^n \frac{\left( \ell_{ij}+\Pi_{ij}\right) \rho^a_{ij}}{x_i-x_j} \und
\bQ_a= \sum_{i=1}^n p_i \bpsi_{i\,a} - \sum_{i \neq j}^n \frac{\left( \ell_{ij}+\Pi_{ij}\right) \brho_{ij\,a}}{x_i-x_j}
\ee
which, together with the Hamiltonian
\be\label{Ham}
H= \frac{1}{2}\sum_{i=1}^n p_i^2 + \frac{1}{2} \sum_{i \neq j}^n \frac{\left( \ell_{ij}+\Pi_{ij}\right)^2 }{\left(x_i-x_j\right)^2}
\ee
indeed obey the $\cN=2M$ super Poincar\'{e} algebra \p{NSP} and thus describe an $\cN=2M$ supersymmetric extension of the $n$-particle Euler--Calogero--Moser model.
To confirm this fact it is most convenient to treat $\Pi_{ij}$ as independent objects, which by themselves span the $so(n)$ algebra~\p{sonPi} and Poisson-commute with the fermions as in~\p{Picom}.
Due to these properties, our construction is valid for an arbitrary number of supersymmetries, in a full analogy
with the extended supersymmetric $su(n)$-spin Calogero model~\cite{SUSYCal}.

\subsection{Superconformal invariance}

The bosonic $n$-particle ECM model admits a dynamical conformal symmetry.
Our $\cN=2 M$ supersymmetric extension with the supercharges \p{QQb} and Hamiltonian \p{Ham} possesses a dynamical superconformal symmetry.
Indeed, starting from the conserved conformal boost current
\be\label{K}
K = \frac{1}{2} \sum_{i=1}^n x_i^2 - t \sum_{i=1}^n x_i p_i +t^2 H,
\ee
the remaining conserved currents can easily be obtained by successively Poisson-commuting the super Poincar\'e generators with $K$.
In this way one finds the full list of conserved currents:
\bea\label{currents}
&& D =  -\frac{1}{2} \sum_{i=1}^n x_i p_i + t H,\qquad \qquad \quad
J^a{}_b= -\sum_{i=1}^n  \psi^a_i \bpsi_{i\,b}- \sum_{i \neq j}^n \rho^a_{ij}\brho_{ij\,b}, \nn \\
&& I^{ab}= - \sum_{i=1}^n  \psi^a_i \psi^b_i- \sum_{i \neq j}^n \rho^a_{ij} \rho^b_{ij},\qquad
\bI_{ab} = \sum_{i=1}^n \bpsi_{i\,a} \bpsi_{i\,b} + \sum_{i \neq j}^n \brho_{ij\,a} \brho_{ij\,b}, \nn \\
&& S^a = \sum_{i=1}^n x_i \psi^a_i - t Q^a, \qquad \qquad \qquad \bS_a = \sum_{i=1}^n x_i \bpsi_{i\,a} - t \bQ_a.
\eea
Together with the supercharges $Q^a, \bQ_a$ \p{QQb}, the Hamiltonian $H$ \p{Ham} and the conformal boost current $K$ \p{K}
they form an $osp(\cN|2)$ superalgebra:
\bea\label{ospN2}
&& \big \{H, K\big \} = 2 D, \quad \big\{H, D \big\} = H, \quad \big\{ K,D\big\} = -K, \nn \\
&& \big\{ J^a{}_b, J^c{}_d\big\} =\im \big( \delta_b^c J^a{}_d -\delta^a_d J^c{}_b\big), \quad
\big\{ J^a{}_b, I^{cd}\big\} = \im \big( \delta_b^c I^{ad}-\delta_b^d I^{ac}\big), \quad
\big\{ J^a{}_b, \bI_{cd}\big\} = -\im \big( \delta^a_c \bI_{bd}-\delta^a_d \bI_{bc}\big),\nn \\
&& \big\{I^{ab}, \bI_{cd}\big\} = \im \big( \delta^a_c J^b{}_d  -\delta^a_d J^b{}_c -\delta^b_c J^a{}_d +\delta^b_d J^a{}_c \big), \nn\\
&& \big\{ D, Q^a\big\} = -\frac{1}{2} Q^a,\quad \big\{ D, \bQ_a\big\} = -\frac{1}{2} \bQ_a, \quad
\big\{ D, S^a\big\} = \frac{1}{2} S^a, \quad \big\{ D, \bS_a\big\} = \frac{1}{2} \bS_a, \nn \\
&& \big\{H, S^a \big\} = - Q^a, \quad \big\{H, \bS_a\big\} = - \bQ_a,\quad \big\{K, Q^a\big\} = S^a, \quad \big\{K, \bQ_a\big\} = \bS_a, \nn \\
&& \big\{ J^a{}_b, Q^c \big\} = \im\, \delta_b^c\, Q^a, \quad \big\{ J^a{}_b, S^c \big\} = \im\, \delta_b^c\, S^a,\quad
\big\{ J^a{}_b, \bQ_c \big\} = -\im\, \delta^a_c\, \bQ_b,\quad \big\{ J^a{}_b, \bS_c \big\} = -\im\, \delta^a_c\, \bS_b, \nn \\
&& \big\{ I^{ab}, \bQ_c \big\} = -\im \big( \delta^a_c Q^b - \delta^b_c Q^a \big ), \quad
\big\{ I^{ab}, \bS_c \big\} = -\im \big( \delta^a_c S^b - \delta^b_c S^a \big), \nn\\
&& \big\{ \bI_{ab}, Q^c \big\} = \im \big( \delta_a^c \bQ_b - \delta_b^c \bQ_a \big), \quad
\big\{ \bI_{ab}, S^c \big\} = \im \big( \delta_a^c \bS_b - \delta_b^c \bS_a \big),  \nn\\
&&  \big\{ Q^a, \bQ_b \big\} = - 2 \im \delta^a_b H, \quad \big\{ S^a, \bS_b \big\} = - 2 \im \delta^a_b K, \nn \\
&& \big\{ Q^a, \bS_b \big\} =  2\, \im\, \delta^a_b\, D +J^a{}_b, \quad \big\{ S^a, \bQ_b \big\} =  2\, \im\, \delta^a_b\, D -J^a{}_b, \nn \\
&&  \big\{ Q^a, S^b \big\} =  I^{ab}, \quad \big\{ \bQ_a, \bS_b\big\} = - \bI_{ab}.
\eea
A $u(M)$ subalgebra is generated by $J^a{}_b$ and extended to an $so(2M)$ subalgebra by adding $I^{ab}$ and $\bI_{ab}$.

\setcounter{equation}{0}
\section{$\cN{=}2$ supersymmetric Euler--Calogero--Moser model in superspace}
With the Hamiltonian description of an $\cN$-extended supersymmetric ECM model at hand, it is quite instructive to construct the superfield
description of the simplest case with $\cN{=}2$ supersymmetry. Such a description may be useful for understanding the general structure of the
given supersymmetric construction, especially the role played by the additional $\rho$-type fermions and the currents $\ell_{ij}$.

To obtain a superspace representation of the $\cN{=}2$ supersymmetric Euler--Calogero--Moser model, defined with $M{=}1$ by the supercharges $Q,\bQ$
\p{QQb} and the Hamiltonian \p{Ham}, one firstly has to solve two tasks:
\begin{itemize}
\item assemble the physical components $x_i, \psi_i, \bpsi_i, \rho_{ij}$ and $\brho_{ij}$ into appropriate $\cN{=}2$ superfields,
\item introduce auxiliary bosonic superfields $v_i, \bv_i$ whose leading components realize $\ell_{ij}$ via bilinear combinations.
\end{itemize}
Let us start with the first task.
{}From the structure of the supercharges $Q, \bQ$ \p{QQb} it is clear that $\cN{=}2$ supersymmetry transforms the coordinates $x_i$ into the
fermions $\psi_i, \bpsi_i$. Thus, one must introduce $n$ bosonic $\cN{=}2$ superfields $\boldsymbol{x}_i$ with the following components,
\be\label{compx}
x_i = \boldsymbol{x}_i|,\quad \psi_i =- \im D \boldsymbol{x}_i|,\quad \bpsi_i = -\im \bD \boldsymbol{x}_i|, \quad A_i = \frac{1}{2} \left[ \bD,D\right] \boldsymbol{x}_i|.
\ee
Here, $|$ denotes the $\theta=\bar{\theta} = 0$ projection,
while $D$ and $\bD$ are $\cN{=}2$ covariant derivatives obeying the relations
\be\label{DDb}
\big\{ D, \bD\big\} = 2 \im \partial_t \und \big\{D, D\big\} = \big\{\bD, \bD\big\}= 0.
\ee

The fermions $\rho_{ij},\; \brho_{ij}$ are put into $n(n{-}1)$ fermionic superfields $\boldsymbol{\rho}_{ij},\; \bar{\boldsymbol{\rho}}_{ij}$,
symmetric and of zero diagonal in the indices $i,j$, i.e.
\be
\boldsymbol{\rho}_{ij}=\boldsymbol{\rho}_{ji}, \quad \bar{\boldsymbol{\rho}}_{ij}=\bar{\boldsymbol{\rho}}_{ji}, \qquad
{\boldsymbol{\rho}}_{ii}=\bar{\boldsymbol{\rho}}_{ii}=0 \quad (\textrm{no sum})\, .
\ee
As $\cN{=}2$ superfields the $\boldsymbol{\rho}_{ij}$ and $\bar{\boldsymbol{\rho}}_{ij}$ contain a lot of components. However, their leading components
$\rho_{ij}$ and $\brho_{ij}$ transform under the $\cN{=}2$ supersymmetry generated by $Q$ and $\bQ$ \p{QQb} as follows,
\bea\label{tr1}
&& \delta_Q  \rho_{ij} \sim  \im \bar\epsilon\, \bigg[ \frac{\psi_i -\psi_j}{x_i-x_j} \rho_{ij} -
\sum_{k \neq i,j}^n \frac{x_i -x_j}{\left( x_i -x_k \right)\left( x_j -x_k\right)} \rho_{ik} \rho_{jk}\bigg], \nn \\
&& \delta_\bQ  \brho_{ij} \sim  \im \epsilon\, \bigg[ \frac{ \bpsi_i -\bpsi_j}{ x_i-x_j} \brho_{ij} -
\sum_{k \neq i,j}^n \frac{ x_i - x_j}{\left( x_i -x_k\right)\left( x_j -x_k\right)} \brho_{ik} \brho_{jk}\bigg].
\eea
To realize these transformations in superspace we are forced to impose the following nonlinear chirality conditions,
\bea\label{chirality1}
&& D \boldsymbol{\rho}_{ij} =  \im \; \bigg[ \frac{\bb{\psi}_i -\bb{\psi}_j}{\bb{x}_i-\bb{x}_j} \bb{\rho}_{ij} -
\sum_{k \neq i,j }^n \frac{\bb{x}_i -\bb{x}_j}{\left(\bb{x}_i -\bb{x}_k\right)\left( \bb{x}_j -\bb{x}_k\right)} \bb{\rho}_{ik} \bb{\rho}_{jk}\bigg], \nn \\
&& \bD \boldsymbol{\brho}_{ij} =  \im \; \bigg[ \frac{\bb{\bpsi}_i -\bb{\bpsi}_j}{\bb{x}_i-\bb{x}_j} \bb{\rho}_{ij} -
\sum_{k \neq i, j}^n \frac{\bb{x}_i -\bb{x}_j}{\left(\bb{x}_i -\bb{x}_k\right)\left( \bb{x}_j -\bb{x}_k\right)} \bb{\brho}_{ik} \bb{\brho}_{jk}\bigg].
\eea
These conditions leave in the superfields $\boldsymbol{\rho}_{ij}$ and $\bar{\boldsymbol{\rho}}_{ij}$ only the components
\be\label{comprho}
\rho_{ij} = \bb{\rho}_{ij}|, \quad B_{ij} =\bD \bb{\rho}_{ij}|, \qquad \brho_{ij} = \bb{\brho}_{ij}|, \quad \bB_{ij} =D \bb{\brho}_{ij}|\;.
\ee

To get the correct Poisson brackets for $\psi_i, \bpsi_i$ and $\rho_{ij}, \brho_{ij}$ \p{PB2} after passing to the Hamiltonian formalism, the kinetic terms
for these fermionic components must read
\be\label{kin1}
\cL^{\psi}_{kin} = \frac{\im}{2} \sum_{i=1}^n \big( \dot{\psi}_i \bpsi_i - \psi_i \dot{\bpsi}_i \big) \und
\cL^{\rho}_{kin} = \frac{\im}{2} \sum_{i,j}^n \big( \dot{\rho}_{ij} \brho_{ij} - \rho_{ij} \dot{\brho}_{ij} \big).
\ee
Altogether, we arrive at the following superfield action for the purely $\cN{=}2$ supersymmetric system with $l_{ij}=0$,
\be\label{action1}
S_0 = \int \diff t\ \diff^2\ \theta \bigg[ -\frac{1}{2} \sum_{i=1}^n D \bb{x}_i \; \bD \bb{x}_i + \frac{1}{2} \sum_{i,j}^n \bb{\rho}_{ij} \bb{\brho}_{ij} \bigg],\qquad
d^2\theta \equiv D \bD.
\ee

Now we come to the second task: realize the $\ell_{ij}$ in terms of auxiliary semi-dynamical variables.
As $so(n)$ generators the $\ell_{ij}$ possess the standard realization
\be\label{L2}
{\hat \ell}_{ij} = \frac{\im}{2} \big( v_i \bv_j - v_j \bv_i\big)
\ee
in terms of $2 n$ bosonic variables $v_i, \bv_i$ subject to
\be\label{vvb}
\big\{ v_i, \bv_j \big\} = -\im \delta_{ij} .
\ee
To implement these new semi-dynamical variables $v_i, \bv_i$ at the superfield level, we have to introduce $2 n$ bosonic superfields $\bb{v}_i, \bb{\bv}_i$.
Additional information about these superfields again comes from the transformation of their first components under $\cN{=}2$ supersymmetry.
These transformations can be learned from the explicit structure of the supercharges $Q,\,\bQ$ \p{QQb}, with the $\ell_{ij}$ being replaced by their realization ${\hat \ell}_{ij}$ \p{L2}:
\be\label{tr2}
\delta_Q  v_i \sim  \im\,\bar\epsilon\;   \sum_{j\neq i}^n \frac{ \rho_{ij} v_j}{ x_i -x_j} \und
\delta_\bQ  \bv_i \sim \im \,  \epsilon\;  \sum_{j \neq i}^n \frac{ \brho_{ij} \bv_j}{ x_i -x_j}.
\ee
This form of the transformations implies that, like $\bb{\rho}_{ij}$ and $\bb{\brho}_{ij}$, also the superfields $\bb{v}_i$ and $\bb{\bv}_i$ are subject to nonlinear chirality conditions,
\be\label{chirality2}
D \bb{v}_i = \im \, \sum_{j\neq i}^n \frac{\bb{\rho}_{ij} \bb{v}_j}{\bb{x}_i -\bb{x}_j} \und
\bD \bb{\bv}_i =  \im \, \sum_{j \neq i}^n \frac{\bb{\brho}_{ij} \bb{\bv}_j}{\bb{x}_i -\bb{x}_j}.
\ee
These conditions leave in the superfields $\bb{v}_i$ and $\bb{\bv}_i$ only the components
\be\label{compv}
v_i = \bb{v}_i|, \quad C_i = -\im \bD \bb{v}_i|, \qquad \bv_i = \bb{\bv}_i|, \quad \bC_i = -\im D \bb{\bv}_i|.
\ee
Finally, to have the brackets \p{vvb}, the kinetic terms for $v_i, \bv_i$ must take the form
\be\label{kin2}
\cL^{v}_{kin} = - \frac{\im}{2} \sum_{i=1}^n \big( \dot{v}_i \bv_i - v_i \dot{\bv}_i \big).
\ee
Therefore, the interaction part ($l_{ij}\neq0$) of the superfield action reads
\be\label{action2}
S_1 = -\frac{1}{2} \int \diff t\ \diff^2 \theta\ \sum_{i=1}^n \bb{v}_i \bb{\bv}_i .
\ee
Combining everything, we conclude that the superfield action should have the form
\be\label{actionF}
S=S_0+S_1 = \int \diff t\ \diff^2 \theta\ \bigg[ -\frac{1}{2} \sum_{i=1}^n D \bb{x}_i \; \bD \bb{x}_i + \frac{1}{2} \sum_{i,j}^n \bb{\rho}_{ij} \bb{\brho}_{ij}
 -\frac{1}{2}  \sum_{i=1}^n \bb{v}_i \bb{\bv}_i \bigg],
\ee
where the superfields $\bb{\rho}_{ij}, \bb{\brho}_{ij}, \bb{v}_i$ and $\bb{\bv}_i$ are subject to the constraints \p{chirality1} and \p{chirality2}, respectively.

Despite the extremely simple form of the superfield action \p{actionF}, its component version looks quite complicated due to the nonlinear chirality constraints
\p{chirality1} and \p{chirality2}. We will write the corresponding component Lagrangian as the sum of a kinetic term $\cL_{kin}$, auxiliary-field terms
$\cL^A_{aux}, \cL^B_{aux}, \cL^C_{aux}$ and a ``matter'' term $\cL_{matter}$,
\be\label{L}
\cL = \cL_{kin}+ \cL^A_{aux}+\cL^B_{aux}+\cL^C_{aux}+\cL_{matter} .
\ee
The explicit form of these terms is
\bea\label{subL}
\cL_{kin}&=& \frac{1}{2} \sum_{i=1}^n {\dot x}_i {\dot x}_i+
\frac{\im}{2} \sum_{i=1}^n \big( \dot{\psi}_i \bpsi_i - \psi_i \dot{\bpsi}_i \big)
+ \frac{\im}{2} \sum_{i,j}^n \big( \dot{\rho}_{ij} \brho_{ij} - \rho_{ij} \dot{\brho}_{ij}\big)
- \frac{\im}{2} \sum_{i=1}^n \big( \dot{v}_i \bv_i - v_i \dot{\bv}_i \big), \nn \\
\cL^A_{aux}&=& \frac{1}{2}\sum_{i=1}^n A_i A_i - \sum_{j\neq i}^n\frac{A_i - A_j}{x_i - x_j}\,\rho_{ij} \brho_{ij} ,\nn \\
\cL^B_{aux}&=& \frac{1}{2}\,\sum_{i,j=1}^n B_{ij}\bB_{ij}
+ \frac{\im}{2}\, \sum_{j\neq i}^n \bigg[ \frac{\psi_i - \psi_j}{x_i - x_j}\, B_{ij} \bar \rho_{ij} +  \frac{\bar\psi_i - \bar\psi_j}{x_i - x_j}\, \bB_{ij} \rho_{ij}
+\frac{B_{ij} v_j \bar v_i}{x_i - x_j} -  \frac{\bB_{ij} v_i \bar v_j}{x_i - x_j} \bigg] \nn \\
&& + \im\,\sum_{k\neq i,j}^n  \frac{x_i - x_j}{(x_i - x_k)(x_j - x_k)}\, \bigg [B_{ik} \rho_{jk} \bar \rho_{ij} +
\bB_{ik}  \bar \rho_{jk} \rho_{ij} \bigg ], \nn\\
\cL^C_{aux} &= & - \frac{1}{2}\,\sum_{i=1}^n C_i \bC_i
+ \frac{1}{2}\,\sum_{j\neq i}^n \frac{1}{x_i-x_j} \bigg[  \rho_{ij} C_j \bar v_i -  \bar\rho_{ij} \bC_j v_i\bigg],\nn \\
\cL_{matter} & = & \frac{1}{2}\,\sum_{i\neq j,k}^n \frac{\rho_{ij} \bar \rho_{ik}}{(x_i-x_j)(x_i-x_k)}v_j \bar v_k
- \frac{1}{2}\, \sum_{j\neq i}^n \bigg[ \frac{\psi_i - \psi_j}{(x_i - x_j)^2}\, \bar\rho_{ij} v_i \bar v_j
-  \frac{\bar\psi_i - \bar\psi_j}{(x_i - x_j)^2}\, \rho_{ij} v_j \bar v_i \bigg] \nn \\
&&+ \frac{1}{2}\sum_{j\neq i}^n \frac{(\psi_i - \psi_j) (\bar\psi_i - \bar\psi_j)}{(x_i-x_j)^2}\, \rho_{ij} \bar \rho_{ij} +
\frac{1}{2} \sum_{i,j \neq k,l} \frac{(x_i-x_j)^2}{(x_i-x_k)(x_j-x_k)(x_i-x_l)(x_j-x_l)}\rho_{ik}\rho_{jk}\brho_{il}\brho_{jl}\nn\\
&&
+ \,\sum_{i,j\neq k}^n \frac{1}{(x_i-x_k)(x_j-x_k)}\bigg [ \frac{x_i-x_j}{x_j-x_k}\left(\psi_j - \psi_k\right)-\left( \psi_i-\psi_j\right)\bigg ]\, \brho_{ik}\brho_{jk}\rho_{ij}  \nn\\
&&
+ \,\sum_{i,j\neq k}^n \frac{1}{(x_i-x_k)(x_j-x_k)}\bigg [ \frac{x_i-x_j}{x_j-x_k}\left(\bpsi_j - \bpsi_k\right)-\left( \bpsi_i-\bpsi_j\right)\bigg ]\, \rho_{ik}\rho_{jk}\brho_{ij}.
\eea
To go on-shell we eliminate the auxiliary fields $A_i, B_{ij}, \bB_{ij}, C_i, \bC_i$ using their equations of motion,
\bea\label{aux3}
&& A_i = 2 \sum_{j\neq i}^n \frac{\rho_{ij} \brho_{ij}}{x_i-x_j}, \qquad C_i = \sum_{j\neq i}^n \frac{\brho_{ij} v_j}{x_i-x_j}, \qquad
 \bC_i = \sum_{j\neq i}^n \frac{\rho_{ij} \bv_j}{x_i-x_j}, \nn \\
&& B_{ij}=\frac{\im}{2} \frac{v_i \bv_j -v_j \bv_i}{x_i-x_j} -\im \frac{\left( \bpsi_i -\bpsi_j\right) \rho_{ij}}{x_i-x_j}+
\im \sum_{k\neq i,j}^n\frac{1}{x_i-x_j} \bigg( \frac{x_i-x_k}{x_k-x_j}\rho_{ik}\brho_{jk} -
\frac{x_j-x_k}{x_k-x_i}\rho_{jk}\brho_{ik} \bigg), \nn \\
&& \bB_{ij}=\frac{\im}{2} \frac{v_i \bv_j -v_j \bv_i}{x_i-x_j} -\im \frac{\left( \psi_i -\psi_j\right) \brho_{ij}}{x_i-x_j}-
\im \sum_{k\neq i,j}^n\frac{1}{x_i-x_j} \bigg( \frac{x_i-x_k}{x_k-x_j}\rho_{jk}\brho_{ik} -
\frac{x_j-x_k}{x_k-x_i}\rho_{ik}\brho_{jk} \bigg).
\eea
After the substitution of the auxiliary components by the expressions~\p{aux3},
a straightforward but slightly tedious calculation brings the Lagrangian~\p{L} to the extremely simple form
\be\label{LN2fin}
\cL = \frac{1}{2} \sum_{i=1}^n {\dot x}_i {\dot x}_i+
\frac{\im}{2} \sum_{i=1}^n \big( \dot{\psi}_i \bpsi_i - \psi_i \dot{\bpsi}_i \big)+ \frac{\im}{2} \sum_{i,j}^n \big( \dot{\rho}_{ij} \brho_{ij} - \rho_{ij} \dot{\brho}_{ij}\big)
- \frac{\im}{2} \sum_{i=1}^n \big( \dot{v}_i \bv_i - v_i \dot{\bv}_i \big) - \sum_{i \neq j}^n \frac{( \hat\ell_{ij}+\Pi_{ij})^2 }{2\left(x_i-x_j\right)^2},
\ee
where $\Pi_{ij}$ is still defined as in \p{Pi} for $a=1$ and ${\hat\ell}_{ij}$ is expressed in terms of semi-dynamical variables as in~\p{L2}.
Thus, the superfield action \p{actionF}, with the superfields $\bb{\rho}_{ij}, \bb{\brho}_{ij}, \bb{v}_i$ and $\bb{\bv}_i$ nonlinearly constrained by \p{chirality1} and \p{chirality2},
indeed describes the $\cN{=}2$ supersymmetric Euler--Calogero--Moser model.

To conclude, let us make a few comments:
\begin{itemize}
\item The nonlinear chirality conditions \p{chirality1} can be slightly simplified by passing to different superfields
$$
\bb{\xi}_{ij}\equiv \frac{\bb{\rho}_{ij}}{\bb{x}_i-\bb{x_j}}\,, \quad \bb{\bar\xi}_{ij}\equiv \frac{\bb{\brho}_{ij}}{\bb{x}_i-\bb{x_j}} \qquad \Rightarrow\qquad
D \bb{\xi}_{ij} + \im \sum_{k=1}^n \bb{\xi}_{ik} \bb{\xi}_{jk}=0, \quad \bD \bb{\bar\xi}_{ij} + \im \sum_{k=1}^n \bb{\bar\xi}_{ik} \bb{\bar\xi}_{jk}=0.
$$
However, the Lagrangian, Hamiltonian and Poisson brackets will look more complicated in terms of $\bb{\xi}_{ij}$ and $\bb{\bar\xi}_{ij}$,
despite the fact that the constraints for these new superfields do no longer involve the superfields $\bb{x}_i$.
\item The auxiliary superfields $\bb{v}_i,\bb{\bv}_i$ cannot be redefined in a similar manner. Thus, the nonlinear chirality constraints \p{chirality2} are unavoidable.
\item The superfield action \p{actionF} looks like a free action for all superfields involved.
However, all interactions are hidden inside the nonlinear chirality constraints \p{chirality1} and \p{chirality2}.
This feature makes our construction quite different from most $\cN{=}2$ supersymmetric mechanics where the interactions are generated via superpotentials.
We are curious whether our mechanism to turn on interactions may be applied elsewhere for constructing new interacting superfield models.
\end{itemize}

\setcounter{equation}{0}
\section{Supersymmetric goldfish model}
To construct an $\cN=2 M$ supersymmetric extension of the bosonic $n$-particle goldfish model \p{eom1} one has to impose a modified version of
the constraints \p{ell0}. It is not too hard to guess such constraints to be
\be\label{susycon}
\widetilde{G}_{ij} \equiv \ell_{ij}+\left(x_i -x_j\right) \sqrt{\dot{x}_i \dot{x_j}}+\Pi_{ij} \approx 0.
\ee
One may check that these constraints weakly commute with the Hamiltonian \p{Ham}, with the supercharges \p{QQb} and with each other, hence they are first class.

To get the equations of motion, one has to evaluate the brackets of all component fields involved with the Hamiltonian
\p{Ham} and then to impose the constraints~\p{susycon}. This results in the following equations of motion:
\bea\label{eom}
&& \dot{x}_i = p_i, \qquad \dot{p}_i = 2 \sum_{j \neq i}^n \frac{p_i\,p_j}{x_i-x_j}, \nn \\
&& \dot{\psi}_i^{\,a} = 2 \sum_{j\neq i}^n \frac{\sqrt{p_i\, p_j}}{x_i-x_j}\, \rho^a_{ij},\qquad
\dot{\bpsi}_{i\,a} = 2 \sum_{j\neq i}^n \frac{\sqrt{p_i\, p_j}}{x_i-x_j}\, \brho_{ij\,a}, \nn \\
&& \dot{\rho}_{ij}^{\,a} = -\frac{\sqrt{p_i\,p_j}}{x_i-x_j} \big( \psi_i^a-\psi_j^a\big)
+ \sum_{k\neq i,j}^n \left[ \frac{\sqrt{p_i\,p_k}}{x_i-x_k}\,\rho^a_{jk}+
\frac{\sqrt{p_j\,p_k}}{x_j-x_k}\,\rho^a_{ik} -2 \delta_{ij} \frac{\sqrt{p_i\,p_k}}{x_i-x_k}\,\rho^a_{ik} \right], \nn \\
&& \dot{\brho}_{ij\,a} = -\frac{\sqrt{p_i\,p_j}}{x_i-x_j} \big( \bpsi_{i\,a}-\bpsi_{j\,a}\big)
+ \sum_{k\neq i,j}^n \left[ \frac{\sqrt{p_i\,p_k}}{x_i-x_k}\,\brho_{jk\,a}+
\frac{\sqrt{p_j\,p_k}}{x_j-x_k}\,\brho_{ik\,a} -2 \delta_{ij} \frac{\sqrt{p_i\,p_k}}{x_i-x_k}\,\brho_{ik\,a} \right].
\eea
The $\cN$-extended supersymmetry transformations, generated by Poisson-commuting
$\im \left(\bar\epsilon_a Q^a+\epsilon^a \bQ_a\right)$  with all components fields and then by imposing the constraints \p{susycon}, have the form
\bea\label{SUSYtr}
\delta x_i &=&\im\, \big( \bar\epsilon_a \psi_i^a + \epsilon^a \bpsi_{i\,a} \big),\quad
\delta p_i = 2 \im\, \sum_{j \neq i}^n \frac{\sqrt{p_i\, p_j}}{x_i-x_j}\big( \bar\epsilon_a \rho^a_{ij}+\epsilon^a \brho_{ij\,a}\big), \nn \\
\delta \psi_i^a &=& 2 \im\, \sum_{j\neq i}^n \frac{\rho^a_{ij}}{x_i-x_j}\big( \bar\epsilon_b \rho^b_{ij}+\epsilon^b \brho_{ij\,b}\big) - \epsilon^a p_i, \quad
\delta \bpsi_{i\,a} = 2 \im\, \sum_{j\neq i}^n \frac{\brho_{ij\,a}}{x_i-x_j}\big( \bar\epsilon_b \rho^b_{ij}+\epsilon^b \brho_{ij\,b}\big) - \bar\epsilon_a p_i, \nn \\
\delta \rho^a_{ij}&=& -\epsilon^a \sqrt{p_i\,p_j}+\epsilon^a \delta_{ij} p_i -\im\, \frac{\psi_i^a -\psi_j^a}{x_i-x_j}
\big(\bar\epsilon_b \rho_{ij}^b + \epsilon^b \brho_{ij\,b}\big)+\im \sum_{k\neq i}^n\frac{\rho^a_{jk}}{x_i-x_k}\big(\bar\epsilon_b \rho^b_{ik}+\epsilon^b \brho_{ik\,b}\big) \nn \\
&+&  \im \sum_{k\neq j}^n\frac{\rho^a_{ik}}{x_j-x_k}\big(\bar\epsilon_b \rho^b_{jk}+\epsilon^b \brho_{jk\,b}\big)
- 2 \im \delta_{ij} \sum_{k\neq i}^n\frac{\rho^a_{ik}}{x_i-x_k}\big(\bar\epsilon_b \rho^b_{ik}+\epsilon^b \brho_{ik\,b}\big), \nn \\
\delta \brho_{ij\,a}&=& -\bar\epsilon_a \sqrt{p_i\,p_j}+\bar\epsilon_a \delta_{ij} p_i -\im\, \frac{\bpsi_{i\,a} -\bpsi_{j\,a}}{x_i-x_j}
\big(\bar\epsilon_b \rho_{ij}^b + \epsilon^b \brho_{ij\,b}\big)
+ \im \sum_{k\neq i}^n\frac{\brho_{jk\,a}}{x_i-x_k}\big(\bar\epsilon_b \rho^b_{ik}+\epsilon^b \brho_{ik\,b}\big)  \nn \\
&+& \im \sum_{k\neq j}^n\frac{\brho_{ik\,a}}{x_j-x_k}\big(\bar\epsilon_b \rho^b_{jk}+\epsilon^b \brho_{jk\,b}\big)
- 2 \im \delta_{ij} \sum_{k\neq i}^n\frac{\brho_{ik\,a}}{x_i-x_k}\big(\bar\epsilon_b \rho^b_{ik}+\epsilon^b \brho_{ik\,b}\big).
\eea
One may verify that these transformations form the $\cN{=}2$ superalgebra and leave the equations of motion \p{eom} invariant.

After imposing the constraints \p{susycon}, the Hamiltonian \p{Ham} and the supercharges \p{QQb} acquire the form
\bea\label{susyHred}
&& H_{red} = \frac{1}{2} \left( \sum p_i \right)^2 \und  \nn \\
&& \left(Q^a\right)_{red} = \sum_i p_i \psi^a_i + \frac{1}{2} \sum_{i \neq j} \sqrt{p_i p_j}\; \rho^a_{ij}, \qquad
\left( \bQ_a\right)_{red} = \sum_i p_i \bpsi_{i\,a} + \frac{1}{2} \sum_{i \neq j} \sqrt{p_i p_j}\; \brho_{ij\,a}.
\eea
It is clear that the correct equations of motion require a deformation of the basic Poisson brackets \p{PB1}, \p{PB2}, similarly to the purely bosonic case \cite{gf}.
We plan to analyze the corresponding deformation of the Poisson brackets elsewhere.

\newpage

\section{Conclusion}
We proposed a novel $\cN$-extended supersymmetric $so(n)$ spin-Calogero model by a direct supersymmetrization of the bosonic Euler--Calogero--Moser system \cite{sCal}.
The constructed model contains
\begin{itemize}
\item $n$ bosonic coordinates $x_i$ which stem from the diagonal part of a real symmetric matrix,
\item the off-shell elements of this symmetric matrix, which enter the supercharges and the Hamiltonian only through $so(n)$ currents $\ell_{ij}$,
\item $\cN\,n$ fermions $\psi^a_i$ and $\bpsi_{i\,a}$, which combine with the $x_i$ to $n$ supermultiplets,
\item $\frac{1}{2}\cN\times n(n{-}1)$ additional fermions $\rho^a_{ij}=\rho^a_{ji}$ and $\brho_{ij \,a} =\left(\rho^a_{ij}\right)^\dagger$ for $i\neq j$.
\end{itemize}
The supercharges $Q^a$ and $\bQ_b$ and the Hamiltonian form an $\cN$-extended Poincar\'{e} superalgebra and have the standard structure up to cubic in the fermions.
Additional conserved currents enlarge this superalgebra to a dynamical $osp(\cN|2)$ superconformal symmetry of the ECM model.
Having performed the Hamiltonian reduction of the ECM model, we obtained the $\cN$-supersymmetric goldfish system for $n$ particles.

The structure of the $so(n)$ spin-Calogero supercharges \p{QQb} and Hamiltonian \p{Ham} is quite similar to the supercharges and the
Hamiltonian of the extended supersymmetric $su(n)$ spin-Calogero model \cite{SUSYCal}. Indeed, the former can be obtained from the latter 
by restricting the $su(n)$ currents $\ell_{ij}$ to the $so(n)$ subalgebra, imposing antisymmetry in their indices, and likewise
restricting the matrix fermions $\rho^a_{ij}$ and $\brho_{ij\, a}$ to be symmetric in their indices.
Upon such a reduction, the composite object $\Pi_{ij}$ also becomes antisymmetric in $(i,j)$ and generates an $so(n)$~algebra.
The first-class constraints $\ell_{ii}+\Pi_{ii}\approx 0$ present in the $su(n)$ spin-Calogero model~\cite{SUSYCal} are then satisfied 
automatically, and the reduced supercharges and Hamiltonian will coincide with the supercharges~\p{QQb} and Hamiltonian~\p{Ham}.
However, the compatibility of this reduction with the extended supersymmetry is {\it not a priori\/} evident and has to be checked explicitly.

The superfield description of our model in the simplest case of $\cN{=}2$ supersymmetry features
\begin{itemize}
\item coordinates $x_i$ and fermions $\psi_i, \bpsi_j$ forming standard unconstrained bosonic superfields of type $(1,2,1)$,
\item fermionic symmetric matrices $\rho_{ij}, \brho_{ij}$ (with vanishing diagonal), subject to nonlinear chirality constraints,
\item $2n$ bosonic $\cN{=}2$ semi-dynamical superfields $v_i, \bv_i$ also obeying some nonlinear chirality constraints.
\end{itemize}
The superspace action contains only the standard kinetic terms for all superfields. It is only the nonlinear constraints which
result in a rather complicated component action. However, after eliminating the auxiliary components via their equations of motion,
the action acquires quite a simple form again, with an interaction quadratic and quartic in the fermions.

The presented $\cN{=}2$ supersymmetric case is not too illuminating, because it can also be constructed without matrix fermions $\rho_{ij}$and $\brho_{ij}$,
in analogy with the $\cN{=}2$ supersymmetric Calogero model~\cite{FM,Wyllard}. One may discard the terms quadratic in $\rho_{ij}$ and $\brho_{ij}$
in the nonlinear chirality constraints \p{chirality1}. Thus, the generic superfield structure of the $\cN$-extended ECM model becomes visible at $\cN{=}4$ only.
We are planning to address this elsewhere.

\vspace{0.5cm}

\noindent{\bf Acknowledgements}\\
We are grateful to V. Gerdt and A. Khvedelidze for stimulating discussions.
This work was partially supported by the Heisenberg-Landau program.
The work of S.K.\ was partially supported by Russian Science Foundation grant 14-11-00598, the one of A.S.\ by RFBR grants 18-02-01046 and 18-52-05002 Arm-a.
This article is based upon work from COST Action MP1405 QSPACE, supported by COST (European Cooperation in Science and Technology).

\end{document}